\documentclass[twocolumn,aps,prl,showpacs,superscriptaddress,amsfonts,amssymb,amsmath]{revtex4}
\usepackage{graphicx}
\begin{document}

\title{Off-diagonal Bethe ansatz and exact solution of a topological spin ring}
\author{Junpeng Cao}
\affiliation{Beijing National Laboratory for Condensed Matter
Physics, Institute of Physics, Chinese Academy of Sciences, Beijing
100190, China}
\author{Wen-Li Yang}
\affiliation{Institute of Modern Physics, Northwest University,
Xi¡¯an 710069, China}
\author{Kangjie Shi}
\affiliation{Institute of Modern Physics, Northwest University,
Xi¡¯an 710069, China}
\author{Yupeng Wang*}
\affiliation{Beijing National Laboratory for Condensed Matter
Physics, Institute of Physics, Chinese Academy of Sciences, Beijing
100190, China}

\begin{abstract}
A general method is proposed for constructing the Bethe ansatz
equations of integrable models without $U(1)$ symmetry. As an
example, the exact spectrum of the $XXZ$ spin ring with M{\" o}bius
like topological boundary condition is derived by constructing a
modified $T-Q$ relation based on the functional connection between
the eigenvalues of the transfer matrix and the quantum determinant
of the monodromy matrix. With the exact solution, the elementary
excitations of the topological $XX$ spin ring is discussed in
detail. It is found that the excitation spectrum indeed shows a
nontrivial topological nature.
\end{abstract}


%75.10.Pq Spin chain models
%03.65.Vf Phases: geometric; dynamic or topological
%71.10.Pm Fermions in reduced dimensions (anyons, composite fermions, Luttinger liquid, etc.)
%02.30.Jr Partial differential equations
%75.10.Jm Quantized spin models

\maketitle

Integrable models play important roles in statistical physics,
quantum field theory and condensed matter physics, because those
models provide some benchmarks for understanding the corresponding
universal classes. Since Yang and Baxter's pioneering works
\cite{yang2,bax1,bax2}, the Yang-Baxter relation has become a
cornerstone for constructing and solving the integrable models.
Especially, the $T-Q$ relation method\cite{bax1,bax2} and the
algebraic Bethe ansatz method\cite{sk,xyz2,yb} developed from the
Yang-Baxter equation have become two very popular methods for
dealing with the exact solutions of the known integrable models.
Generally speaking, there are two classes of integrable models. One
possesses $U(1)$ symmetry and the other does not. Three well known
examples without $U(1)$ symmetry are the $XYZ$ spin chain
\cite{xyz2,xyz1}, the $XXZ$ spin chain with antiperiodic boundary
condition \cite{G1,G2,G3,G4,G5,G6,G7} and the ones with unparallel
boundary fields \cite{G8,G9,G10,G11,G12}. It has been demonstrated
that the algebraic Bethe ansatz and $T-Q$ relation can successfully
diagonalize the integrable models with $U(1)$ symmetry. However, for
those without $U(1)$ symmetry, only some very special cases such as
the $XYZ$ spin chain with even site number\cite{xyz2,xyz1} and the
$XXZ$ spin chain with constrained unparallel boundary fields
\cite{G8,G9,G10} can be dealt with because of the existence of a
proper ``local vacuum state" in these special cases. The main
obstacle applying the algebraic Bethe ansatz and Baxter's method to
general integrable models without $U(1)$ symmetry lies in the
absence of such a ``local vacuum". A promising method for
approaching such kind of problems is Sklyanin's separation of
variables method\cite{sk1,sk2} which has been recently applied to
some integrable models\cite{G4,G5,G6,G7,G11,G12}. However, a
systematic method is still absent to derive the usual Bethe ansatz
equations (BAEs) which are crucial for studying the physical
properties in the thermodynamic limit.

In this letter, we develop a general method for dealing with the
integrable models without $U(1)$ symmetry. The central point lies in
how to construct a $T-Q$ relation and the usual BAEs for those
models based on the connection between two basic invariants of the
monodromy matrix, i.e., its trace (transfer matrix) and its quantum
determinant which do not depend on the basis choice and whether
there exists a reference state. As a concrete example, we study the
spectrum of the $XXZ$ spin ring with M{\" o}bius like topological
boundary condition, as it is tightly related to the recent study on
the topological states of matter. In fact, the topological boundary
problem in many body systems has been rarely touched. With the
inhomogeneous $XXZ$ topological spin ring model, we elucidate how
our method works to derive the exact spectrum  and the BAEs by
constructing and solving a recursive functional equations.
Particular attention is focused on the elementary excitations of the
homogeneous $XX$ spin ring with antiperiodic boundary condition, as
it is the simplest quantum realization of the M{\" o}bius stripe.
Our exact solution shows that the elementary excitations of this
simple model indeed exhibit a nontrivial topological nature.

We start from the following model Hamiltonian
\begin{eqnarray}
H=-\sum_{j=1}^{N}\left[\sigma_{j}^{x}\sigma_{j+1}^{x}+\sigma_{j}^{y}
\sigma_{j+1}^{y}+\cosh\eta\sigma_{j}^{z} \sigma_{j+1}^{z}\right],
\label{hami}
\end{eqnarray}
with the anti-periodic boundary conditions
$\sigma_{N+1}^\alpha=\sigma_{1}^x\sigma_{1}^\alpha\sigma_{1}^x$. $N$
is the site number of the system and $\sigma_{j}^{\alpha}$
$(\alpha=x, y, z)$ is the Pauli matrix on the site $j$ along the
$\alpha$ direction. With such a topological boundary condition, the
spin on the $N$-th site connects with that on the first site after
rotating $\pi$-angle along the $x$-direction (a kink on the $(N,1)$
bond) and forms a torus in the spin space. With an unitary
transformation $U_nHU_n^{-1}$, $U_n=\prod_{j=1}^n\sigma_j^x$, the
kink can be shifted to the $(n,n+1)$ bond without changing the
spectrum of the Hamiltonian. Notice here the braiding is in the
quantum space rather than in the real space and therefore the
present model describes a quantum M{\" o}bius stripe. We define a
$Z_2$ operator ${U}_N=\prod_{j=1}^N\sigma^x_j$. It can be easily
checked that ${U_N}^2=1$ and $[H,U_N]=0$. Therefore, the present
model possesses a global $Z_2$ invariance, indicating the double
degeneracy of the eigenstates.

The integrability of the present model is associated with the
following Lax operator
\begin{eqnarray*}
L_{0j}(\lambda)=\left(\begin{array}{cc}
\sinh({\tilde\lambda}_j+\frac{\eta}{2}(1+\sigma^z_j)) &\sinh\eta \sigma_j^-\\
\sinh\eta \sigma_j^+ &
\sinh({\tilde\lambda}_j+\frac{\eta}{2}(1-\sigma^z_j))
  \end{array}\right), \label{hl-1}
\end{eqnarray*}
and the monodromy matrix
\begin{eqnarray*}
T_0(\lambda) = L_{01}(\lambda)\cdots
L_{0N}(\lambda)=\left(\begin{array}{cc} A(\lambda) & B(\lambda) \\
C(\lambda)& D(\lambda)
  \end{array}\right),
\end{eqnarray*}
where ${\tilde\lambda}_j=\lambda-\theta_j$, $\lambda$ is the
spectral parameter and $\theta_j$ are the site inhomogeneous
constants; $\eta$ is the crossing parameter as usual; the index $0$
indicates the auxiliary space and $j$ indicates the quantum space.
Both the Lax operator and the monodromy matrix satisfy the
Yang-Baxter relation
\begin{eqnarray}
&&
R_{12}(\lambda_1-\lambda_2)L_{1j}(\lambda_1)L_{2j}(\lambda_2)\nonumber
\\ && \qquad
=L_{2j}(\lambda_2)L_{1j}(\lambda_1)R_{12}(\lambda_1-\lambda_2),\nonumber\\
&& R_{12}(\lambda_1-\lambda_2)T_1(\lambda_1)T_2(\lambda_2)\nonumber
\\ && \qquad =T_2(\lambda_2)T_1(\lambda_1)R_{12}(\lambda_1-\lambda_2), \label{tybr}
\end{eqnarray}
with $R_{12}(\lambda)=L_{12}(\lambda|\theta_j=0)$. The transfer
matrix of the system is defined as
\begin{equation}
\tau(\lambda)=tr_0 \sigma_0^x T_0(\lambda)=B(\lambda)+C(\lambda),
\label{transfer}
\end{equation}
where $tr_0$ means tracing the auxiliary space. From
Eq.(\ref{tybr}), one can prove that the transfer matrices with
different spectral parameters are mutually commutative, i.e.,
$[\tau(\lambda), \tau(\mu)]=0$. Therefore, $\tau(\lambda)$ serves as
the generating functional of the conserved quantities of the
corresponding system. The first order derivative of logarithm of the
transfer matrix gives the Hamiltonian (\ref{hami})
\begin{equation}
H=-2{\sinh\eta}\frac{\partial \ln \tau(\lambda)}{\partial
\lambda}|_{\lambda=0,\theta_j=0}+N\cosh\eta. \label{h}
\end{equation}

Define the state $|0\rangle=\otimes |\uparrow\rangle_j$. From the
definition of the Lax operator we obtain
\begin{eqnarray}
&& C(\lambda)|0\rangle=0,{~~~}A(\lambda)|0\rangle=a(\lambda)|0\rangle, \nonumber \\
&& D(\lambda)|0\rangle=d(\lambda)|0\rangle,
\end{eqnarray}
where $a(\lambda)=\prod_{n=1}^N\sinh(\lambda-\theta_n+\eta)$ and
$d(\lambda)=\prod_{n=1}^N\sinh(\lambda-\theta_n)$. Before going
further, we introduce the following useful formula\cite{yb}
\begin{eqnarray}
&&
C(\lambda)\prod_{l=1}^nB(\mu_l)|0\rangle=\sum_{l=1}^nM_n^l(\lambda,\{\mu_j\})B_{n-1}^l|0\rangle
\nonumber \\  &&\qquad\qquad  +\sum_{k>l}{\tilde
M}_n^{kl}(\lambda,\{\mu_j\}) B_{n-1}^{kl}|0\rangle,
\end{eqnarray}
which can be obtained from the commutation relations derived from
the Yang-Baxter relation (\ref{tybr}), where
\begin{eqnarray}
B_{n-1}^l=\prod_{j\neq l}^nB(\mu_j),\quad  B_{n-1}^{kl}=
B(\lambda)\prod_{j\neq k,l}^n B(\mu_j)\nonumber
\end{eqnarray}
and
\begin{eqnarray}
&&M_n^l(\lambda,\{\mu_j\})=g(\lambda,\mu_l)a(\lambda)d(\mu_l)\prod_{j\neq
l}f(\lambda,\mu_j)f(\mu_j,\mu_l)\nonumber\\
&& \qquad +g(\mu_l,\lambda)a(\mu_l)d(\lambda)\prod_{j\neq l}f(\mu_j,\lambda)f(\mu_l,\mu_j),\\
&& {\tilde M}_{n}^{kl}(\lambda,\{\mu_j\})=
g(\lambda,\mu_k)g(\mu_l,\lambda)f(\mu_l,\mu_k)a(\mu_l)d(\mu_k)\nonumber\\
&& \qquad \times \prod_{j\neq
k,l}f(\mu_j,\mu_k)f(\mu_l,\mu_j)\nonumber\\
&& \qquad +g(\lambda,\mu_l)g(\mu_k,\lambda)f(\mu_k,\mu_l)a(\mu_k)d(\mu_l)\nonumber\\
&&\qquad \times \prod_{j\neq k,l}f(\mu_j,\mu_l)f(\mu_k,\mu_j),\\
&&g(\lambda,\mu)=\frac{\sinh\eta}{\sinh(\mu-\lambda)},\quad
f(\lambda,\mu)=\frac{\sinh(\lambda-\mu-\eta)}{\sinh(\lambda-\mu)}.
\nonumber
\end{eqnarray}
We adopt the procedure introduced in\cite{G3}. Suppose
$|\Psi\rangle$ is an eigenstate of $\tau(\lambda)$ and independent
of $\lambda$. We have
$\tau(\lambda)|\Psi\rangle=\Lambda(\lambda)|\Psi\rangle$. In
addition, we define $F_n(\{\mu_j\})=\langle\Psi|\prod_{j=1}^n
B(\mu_j)|0\rangle$ and put $F_0=\langle\Psi|0\rangle=1$. Consider
the quantity $\langle\Psi|\tau(\lambda)\prod_{j=1}^n
B(\mu_j)|0\rangle$. By acting $\tau(\lambda)$ right and left
alternatively, we have the following functional relations
\begin{eqnarray}
&& \Lambda(\lambda)F_n=\sum_l M_n^l(\lambda)
F_{n-1}^l+\sum_{k>l}{\tilde
M}_n^{kl}(\lambda)F_{n-1}^{kl}+F_{n+1},\nonumber \\
&& F_1(\lambda)=\Lambda(\lambda), \label{re} \\
&& F_{N+1}\equiv 0, \nonumber
\end{eqnarray}
where $F_n=F_n(\{\mu_j\})$, $F_{n-1}^l = F_{n-1}(\{\mu_j\}_{j\neq
l})$, $F_{n-1}^{kl} = F_{n-1} (\lambda, \{\mu_j\}_{j\neq k,l})$ and
$\{\mu_j\}$ indicating the parameter set $\{\mu_1,\cdots,\mu_n\}$
for $n=1, \cdots, N$. Notice that we have $N+2$ equations and $N+2$
unknown functions $\Lambda$ and $F_n$. The function $F_n(\{\mu_j\})$
is symmetric by exchanging the variables $\mu_j$ because of
$[B(\mu_j),B(\mu_l)]=0$ and is a degree $N-1$ trigonometrical polynomial. The eigenvalue $\Lambda(\lambda)$ therefore can
be parameterized as
\begin{eqnarray}
\Lambda(\lambda)=\Lambda_0\prod_{j=1}^{N-1}e^{z_j}\sinh(\lambda-z_j),\label{Ploy-1}
\end{eqnarray}
where $\Lambda_0$ is a constant and $\{z_1,\cdots,z_{N-1}\}$ is a
set of roots of $\Lambda(\lambda)$ with $\Lambda(z_j)=0$. The
recursion equations (\ref{re}) determine the eigenvalue $\Lambda(\lambda)$.
From Eq.(\ref{h}) we can easily derive the eigenvalue of the Hamiltonian
as
\begin{eqnarray}
E&=&-2{\sinh\eta}\frac{\partial \ln \Lambda(\lambda)}{\partial
\lambda}|_{\lambda=0,\theta_j=0}+N\cosh\eta\nonumber\\
&=&-2{\sinh\eta}\sum_{j=1}^{N-1}\coth z_j+N\cosh\eta.\label{Eigenvalues}
\end{eqnarray}

Since $d(\theta_j)=0$, all the functions $M_n^j$ and ${\tilde
M}_n^{jk}$ are zero as long as their variables belong to the
parameter set $\{\theta_1,\cdots,\theta_N\}$ and
$\theta_j\neq\theta_k\neq\theta_l\pm\eta$. Therefore, the following
relations hold
\begin{eqnarray}
F_n(\theta_1,\cdots,\theta_n)=\prod_{j=1}^n\Lambda(\theta_j).
\end{eqnarray}
From the $n=N$ case of Eq.(\ref{re}) we obtain
\begin{eqnarray}
\Lambda(\lambda)&=&\sum_{j=1}^N\frac{a(\theta_j)d(\lambda)}{\Lambda(\theta_j)}g(\theta_j,\lambda)\prod_{l\neq
j}^Nf(\theta_j,\theta_l)f(\theta_l,\lambda)\nonumber\\
&=&-\sum_{j=1}^N\frac{a(\theta_j)d(\theta_j-\eta)}{\Lambda(\theta_j)d_j(\theta_j)\sinh(\lambda-\theta_j+\eta)}a(\lambda),
\label{BAE-1}
\end{eqnarray}
with $d_j(\theta_j)=\prod_{l\neq j}^N\sinh(\theta_j-\theta_l)$. This
equation gives the closed recursive solution of $\Lambda(\lambda)$.
Putting $\lambda\to\theta_j-\eta$, we readily have
\begin{eqnarray}
&&\Lambda(\theta_j)\Lambda(\theta_j-\eta)=\Delta_q(\theta_j),{~~~}j=1,\cdots,N,\label{BAE-n}
\end{eqnarray}
where $\Delta_q(\theta_j)=-a(\theta_j)d(\theta_j-\eta)$ is the
quantum determinant\cite{yb}. Similar relations were also derived
in\cite{G6,G7,G11} with the separation of variables method. The
above equations determine the $N-1$ roots $\{z_j\}$ and $\Lambda_0$
in Eq.(\ref{Ploy-1}). In fact, the operator identity
$B(\theta_j)B(\theta_j-\eta)=0$ can be demonstrated with the
definition of the monodromy matrix. With this operator identity and
considering the quantity
$\langle\Psi|\tau(\theta_j)\tau(\theta_j-\eta)|0\rangle$, one can
easily deduce Eq.(\ref{BAE-n}). Taking the limit of Eq.(\ref{BAE-n})
with $\theta_j\rightarrow 0$ leads to the following equations which
completely determine the spectrum $\Lambda(\lambda)$ of the
homogeneous model
\begin{eqnarray}
&&\frac{\partial^l}{\partial u^l}\ln (-\sinh^N(u+\eta)\sinh^N(u-\eta))\left|\right._{u=0}\nonumber\\
&&=
\frac{\partial^l}{\partial u^l}\ln(\Lambda(u)\Lambda(u\hspace{-0.08truecm}-\hspace{-0.08truecm}
\eta))\left|\right._{u=0},\,
l=0,\ldots,N\hspace{-0.08truecm}-\hspace{-0.08truecm}1.\label{BAE-2}
\end{eqnarray}
%Alternatively,
%\begin{eqnarray}
%\Lambda(\lambda)=-\sum_{j=1}^N\frac{a(\theta_j)d(\theta_j-\eta)}{\Lambda(\theta_j)d_j(\theta_j)\sinh(\lambda-\theta_j)}d(\lambda).
%\end{eqnarray}
%With Eq.(9),(13) and (15) we have the Bethe ansatz equations about
%$z_n$ as
%\begin{eqnarray}
%\frac{{\tilde a}(z_n)}{{\tilde
%d}(z_n)}=\prod_{l=1}^{N-1}\frac{\sinh(z_n-z_l+\eta)}{\sinh(z_n-z_l-\eta)}.
%\end{eqnarray}
However, these relations are quite hard to be used to study the
physical properties, especially in the thermodynamic limit. Thus a
proper set of BAEs in the usual form is still crucial. As
$\Lambda(\lambda)$ is a trigonometrical polynomial of degree
$N\hspace{-0.08truecm}-\hspace{-0.08truecm}1$ with the very
periodicity $\Lambda(\lambda+i\pi)=(-1)^{N-1}\Lambda(\lambda)$, we
conjecture the following modified $T-Q$ relation\cite{bax1,bax2}
\begin{eqnarray}
\Lambda(\lambda)&&=e^{\lambda}a(\lambda)\frac{Q_1(\lambda-\eta)}{Q_2(\lambda)}
-e^{-\lambda-\eta}d(\lambda)\frac{Q_2(\lambda+\eta)}{Q_1(\lambda)}\nonumber\\
&&-b(\lambda)\frac{a(\lambda)d(\lambda)}{Q_1(\lambda)Q_2(\lambda)},
\label{tqre}
\end{eqnarray}
where
\begin{eqnarray}
Q_1(\lambda)&&=\prod_{j=1}^{M}\sinh(\lambda-\mu_j), \nonumber \\
Q_2(\lambda)&&=\prod_{j=1}^{M}\sinh(\lambda-\nu_j),
\end{eqnarray}
and $b(\lambda)$ is an adjust function. For $N$ even, $M=\frac N2$,
\begin{eqnarray}
b(\lambda)=e^{i\phi_1+\lambda}-e^{i\phi_2-\lambda-\eta},
\end{eqnarray}
with
\begin{eqnarray}
i\phi_1&=&\sum_{j=1}^N\theta_j-M\eta-2\sum_{j=1}^M\mu_j,\nonumber\\
-i\phi_2&=&\sum_{j=1}^N\theta_j-M\eta-2\sum_{j=1}^M\nu_j,\label{fi}
\end{eqnarray}
to cancel the leading terms in Eq.(\ref{tqre}) when
$\lambda\to\pm\infty$. Obviously, the conjectured $\Lambda(\lambda)$
satisfies Eq.(\ref{BAE-n}) automatically. The BAEs determined by the
regularity of $\Lambda(\lambda)$ (which ensures $\Lambda(\lambda)$
to be a trigonometrical polynomial of degree $N-1$) read
\begin{eqnarray}
&&d(\nu_j)=\frac{e^{\nu_j}}{b(\nu_j)}Q_1(\nu_j-\eta)Q_1(\nu_j),\nonumber\\
&&a(\mu_j)=-\frac{e^{-\mu_j-\eta}}{b(\mu_j)}Q_2(\mu_j+\eta)Q_2(\mu_j),\label{BAE-two}\\
&& \qquad\qquad \qquad j=1,\cdots,\frac N2.\nonumber
\end{eqnarray}
%\begin{eqnarray}
%d(\nu_j)=\frac 1{e^{i\phi_1}-e^{i\phi_2-\eta-2\nu_j}}Q_1(\nu_j-\eta)Q_1(\nu_j),\nonumber\\
%a(\mu_j)=\frac1{e^{i\phi_2}-e^{i\phi_1+2\mu_j+\eta}}Q_2(\mu_j+\eta)Q_2(\mu_j),\label{BAE-two}\\
%j=1,\cdots,\frac N2.\nonumber
%\end{eqnarray}
The BAEs for the homogeneous model are exactly the above equations
by putting all $\theta_j=0$.
%For real $\eta$, we put $\mu_j\to
%i\bar\mu_j-\frac \eta2$, $\nu_j\to i\bar\nu_j-\frac\eta2$. From
%Eqs.(\ref{fi}) and (\ref{BAE-two}) we have $\phi_1=-\phi_2^*$,
%$\bar\mu_j=\bar\nu_j^*$ and the BAEs can be further reduced to
%\begin{eqnarray}
%&&(e^{-i\phi_1^*+2i\bar\mu_j}-e^{i\phi_1})\sin^N(\bar\mu_j+i\frac\eta2)
%\nonumber\\
%&&\quad =\prod_{l=1}^M\sin(\bar\mu_j-\bar\mu_l^*+i\eta)\sin(\bar\mu_j-\bar\mu_l^*),\\
%&&\qquad \qquad \phi_1=-2\sum_{j=1}^M\bar\mu_j,{~~} M=\frac
%N2.\nonumber
%\end{eqnarray}
The eigenvalues of Hamiltonian (\ref{hami}) take the following form
\begin{eqnarray}
&&E(\{\mu _j,\nu_j\})=
2\sinh\eta\sum_{j=1}^{M}\left\{\frac{\cosh(\mu_j+\eta)}{\sinh(\mu_j+\eta)}\right.\nonumber\\
&&\left.-\frac{\cosh(\nu_j)}{\sinh(\nu_j)}\right\}
+N\cosh\eta-2\sinh\eta.\label{ei2}
\end{eqnarray}
For odd $N$, we put $M=(N+1)/2$ and
\begin{eqnarray}
b(\lambda)=\frac{1}{2}[e^{i\phi_1+2\lambda}+e^{i\phi_2-2\lambda-2\eta}],
\label{bb}
\end{eqnarray}
where $\phi_1$ and $\phi_2$ take the same form as Eq. (\ref{fi})
with $M=(N+1)/2$ and $\theta_j=0$.
%\begin{eqnarray}
%d(\nu_j)=\frac 2{e^{i\phi_1+\nu_j}-e^{i\phi_2-2\eta-3\nu_j}}Q_1(\nu_j-\eta)Q_1(\nu_j),\nonumber\\
%a(\mu_j)=\frac 2{e^{i\phi_2-\mu_j-\eta}-e^{i\phi_1+3\mu_j+\eta}}Q_2(\mu_j+\eta)Q_2(\mu_j),\label{BAE-two2}\\
%j=1,\cdots,\frac {N+1}{2}.\nonumber
%\end{eqnarray}
In this case, the BAEs and the eigenvalue of the Hamiltonian are
still given by Eq.(\ref{BAE-two}) and Eq.(\ref{ei2}), respectively.
The nested nature of the BAEs is due to the topological boundary and
broken $U(1)$ symmetry.

Generally, the Bethe roots distribute in the whole complex plane
with the selection rules $\mu_j\neq \mu_l$, $\mu_j\neq\nu_l$ and
$\mu_j\neq\nu_l-\eta$ which ensure the simplicity of ``poles" in our
$T-Q$ ansatz. Numerical solutions of the BAEs for small size (up to
$N=6$) with random choice of $\eta$ indicate that the BAEs indeed give
the complete solutions of the model (namely, the eigenvalues calculated
from the BAEs coincide exactly to those obtained from exact
diagonalization). Numerical results for $N=3$ and $\eta=\ln2$ are
shown in TABEL I.
\begin{table*}[h]
\caption{Numerical solutions of the BAEs for $\eta=\ln2$, $N=3$ and
$M=2$. $E_0$ is the eigenenergy and $elv$ indicates the number of
the energy levels. The eigenvalues are exactly the same to those of
the exact diagonalization}.

\begin{tabular}{rrrr|rr}
\hline\hline$\mu_{1} $&$\mu_{2} $&$\nu_{1} $&$\nu_{2} $&$E_0$ & elv\\
\hline
 $-0.82276-0.50000{\rm i}\pi $ & $0.64639-0.50000{\rm i}\pi $ & $-1.33954-0.50000{\rm i}\pi $ & $0.12962-0.50000{\rm i}\pi $ &$-3.02200$ &$1$\\[0pt]
 $-0.48493-0.26054{\rm i}\pi $ & $-0.48493+0.26054{\rm i}\pi $ & $-0.20821-0.26054{\rm i}\pi $ & $-0.20821+0.26054{\rm i}\pi $ &$-3.02200$ &$1$\\[0pt]
 $-1.10839+0.19525{\rm i}\pi $ & $-0.21237+0.00207{\rm i}\pi $ & $-0.48078+0.00207{\rm i}\pi $ & $0.41524+0.19525{\rm i}\pi $ &$-1.25000$ &$2$\\[0pt]
 $-0.24763-0.20157{\rm i}\pi $ & $-0.23124-0.49575{\rm i}\pi $ & $-0.46190-0.49575{\rm i}\pi $ & $-0.44551-0.20157{\rm i}\pi $ &$-1.25000$ &$2$\\[0pt]
 $-0.24763+0.20157{\rm i}\pi $ & $-0.23124+0.49575{\rm i}\pi $ & $-0.46190+0.49575{\rm i}\pi $ & $-0.44551+0.20157{\rm i}\pi $ &$-1.25000$ &$2$\\[0pt]
 $-1.10839-0.19525{\rm i}\pi $ & $-0.21237-0.00207{\rm i}\pi $ & $-0.48078-0.00207{\rm i}\pi $ & $0.41524-0.19525{\rm i}\pi $ &$-1.25000$ &$2$\\[0pt]
 $-0.35506+0.00000{\rm i}\pi $ & $-0.11541+0.50000{\rm i}\pi $ & $-0.57773-0.50000{\rm i}\pi $ & $-0.33809+0.00000{\rm i}\pi $ &$5.52200$ &$3$\\[0pt]
 $-0.44482-0.09194{\rm i}\pi $ & $-0.44482+0.09194{\rm i}\pi $ & $-0.24833-0.09194{\rm i}\pi $ & $-0.24833+0.09194{\rm i}\pi $ &$5.52200$ &$3$\\
\hline\hline
\end{tabular}
\end{table*}
For imaginary $\eta$, the numerical simulations for  $N=8,10$
indicate that the distribution of the Bethe roots in the ground
state is almost on a straight line $Im\{\mu_j\}\sim Im\{\nu_j\}\sim
\frac{i\pi-\eta}2\equiv-\frac{\bar\eta}2$ and $\{Re \mu_j\}\sim\{-Re
\nu_j\}$. This strongly suggests that in the thermodynamic limit
$N\to\infty$, the BAEs for the ground state can be rewritten as
(Eq.(20) over its complex conjugate)
\begin{eqnarray}
\frac{\sinh^N({\bar\mu}_j-\frac{\bar\eta}2)}{\sinh^N({\bar\mu}_j+\frac{\bar\eta}2)}=e^{i\chi_j}\prod_{l=1}^M\frac{\sinh{\bar\mu}_j+{\bar\mu}_l-\bar\eta)}
{\sinh({\bar\mu}_j+{\bar\mu}_l+\bar\eta)},\label{BAE-G}
\end{eqnarray}
where ${\bar\mu}_j=Re(\mu_j)$ and $\chi_j$ accounts for the small
deviation of $\mu_j$ from ${\bar\mu}_j-\frac{\bar\eta}2$. Based on
the above equation, the energy can be derived with the ordinary
method\cite{yb}. $\{\chi_j\}$ contribute a finite boundary energy
and $N^{-1}$ small correction to the excitations. To get the
excitations, we need either exchange two modes ${\bar\mu}_j$ and
${\bar\nu}_l=-{\bar\mu}_j$ or put them onto the complex plane
because of the selection rules, indicating the topological nature of
the system.

To show the physical effect of the topological boundary clearly, let
us focus on the $\eta=i\pi/2$ case, i.e., the topological $XX$ spin
ring or equivalently the topological free fermion ring (via a
Jordan-Wigner transformation). This model is tightly related to the
problem of a Josephson junction embedded in a Luttinger liquid (see,
for example, Ref.\cite {add1,add2,add3}). We note the BAEs about the
roots $\{z_j\}$ can also be derived from the recursive equations
Eq.(\ref{re}).
%In the homogeneous case, the roots $\{z_j\}$ satisfy
%the equation $a^2(z_j)=d^2(z_j)$ (see below). This equation has
%$2(N-1)$ solutions. However, we need only $N-1$ roots to form a
%solution set. That means a selection rule is needed.
%\begin{eqnarray}
%M^l_n(\lambda)&=&\frac{i[a(\lambda)d(\mu_l)-a(\mu_l)d(\lambda)]}{\sinh(\mu_l-\lambda)}\nonumber\\
%&\times&\prod_{j\neq l}\coth(\mu_j-\mu_l)\coth(\lambda-\mu_j), \\
%{\tilde M}^{kl}_n(\lambda)&=&\frac{i[a(\mu_l)d(\mu_k)-a(\mu_k)d(\mu_l)]}{\sinh(\lambda-\mu_k)\sinh(\lambda-\mu_l)\tanh(\mu_k-\mu_l)}\nonumber\\
%&\times&\prod_{j\neq k,l}\coth(\mu_j-\mu_k)\coth(\mu_l-\mu_j).
%\end{eqnarray}
From the $n=N$ case of Eq.(\ref{re}) we obtain that
\begin{eqnarray}
F_N(z,\{\mu_l\})=\frac{1}{\Lambda_0}\sum_{j=1,z}^{N-1}{M^\prime}^j_{N}F_{N-1},\label{mu1}
\end{eqnarray}
where $\mu_1, \cdots, \mu_{N-1}=\{\mu_l\}$ are free parameters; $z$
is one of the roots of the eigenvalue $\Lambda(\lambda)$;
${M^\prime}^j_{N}=\lim_{\mu\to\infty}e^{-(N-1)\mu}M_N^j(\mu,z,\{\mu_l\})$.
Meanwhile, $\Lambda(z)$ acting on $F_{N-1}(\{\mu_l\})$ gives
\begin{eqnarray}
&&F_N(z, \{\mu_l\})=-\sum_{j=1}^{N-1}M^j_{N-1}(z, \{\mu_l\})
F^j_{N-2}\nonumber  \\&&\qquad\qquad\qquad  -{\tilde
M}^{jk}_{N-1}(z,\{\mu_l\})F^{jk}_{N-2}(z, \{\mu_l\}). \label{mu2}
\end{eqnarray}
From Eqs.(\ref{mu1}) and (\ref{mu2}), we have
\begin{eqnarray}
&&F^j_{N-2}=-\frac{1}{M^j_{N-1}(z, \mu_j)}\left[\sum_l
{M^\prime}^l_{N}F_{N-1}\right.\nonumber \\
&&\qquad \left.+\sum_{l\neq j} M^l_{N-1}F^l_{N-2}-\sum_{k>l} {\tilde
M}^{kl}_{N-1}F^{kl}_{N-2}\right]. \label{mu3}
\end{eqnarray}
\begin{figure}
\includegraphics[width=8cm]{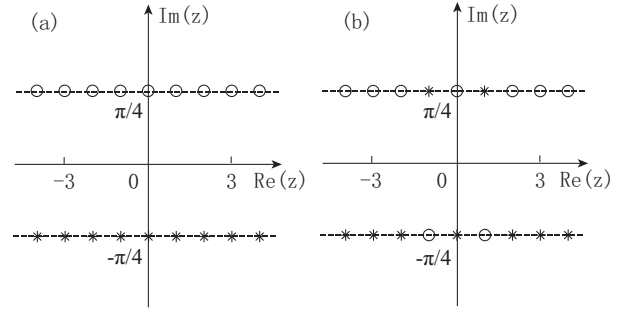}
\caption{(a) Schematic diagram of the ground state of the $XX$
topological spin ring. The states in the lower solution line are all
filled and the upper solution line is unoccupied.(b) The elementary
excitation of the $XX$ topological spin ring. The ``particle" in the
upper solution line must correspond to a ``hole" in the lower
solution line with exactly the same real part.}
\end{figure}
\begin{figure}
\includegraphics[width=6cm]{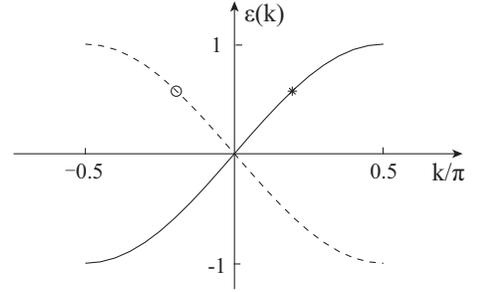}
\caption{ Schematic diagram of the excitation spectrum of the $XX$
topological spin ring. The ``particle" and ``hole" carry exactly the
same energy and opposite momenta.}
\end{figure}
The singular point of Eq.(\ref{mu3}) is $\mu_j=z\pm i\pi/2$. Since
$F_{N-1}$ is a polynomial for all the variables $\mu_j$, the residue
of the right hand side of Eq.(\ref{mu3}) must be zero at the
singular point. Notice the fact that $M_{n}^j(z,\{\mu_j\})=0$ if
$\mu_j\neq z\pm i\pi/2$ and ${\tilde M}^{kl}_{n}(z,\{\mu_j\})=0$ as
long as $\mu_k$ or $\mu_l=z\pm i\pi/2$. We readily have the
conclusion that $F_{N-1}(z,\{\mu_l\})$ is proportional to
$F_{2}(z,z\pm i\pi/2)$ when one of $\mu_l$'s is equal to $z\pm
i\pi/2$, which must be zero according to the above analysis. This
gives the constrain condition of the root $z$ as
$a(z)d(z+i\pi/2)=d(z)a(z+i\pi/2)$. Therefore, the roots $z_n$ of
$\Lambda(\lambda)$ satisfy the following Bethe ansatz equation
\begin{eqnarray}
\coth^{2N}(z_n)=1, \qquad z_j \neq z_k \pm \frac\pi 2i.\label{bea}
\end{eqnarray}
Equivalently, we have
\begin{eqnarray}
\coth(z_n)=e^{\frac {i\pi n}{N}}\equiv e^{ik_n}, \,\,n=
\pm1,\cdots,\pm(N-1).
\end{eqnarray}
The $N-1$ pair solutions $\{z_j, z_j+\frac\pi 2i\}$ $mod (i\pi)$ are
located on two lines with imaginary part $\pm i\pi/4$. The root sets
are formed by choosing one and only one in each pair. This selection
rule comes from that the poles of the right-hand-side of
Eq.(\ref{mu3}) do not enter into the set of roots $\{z_j\}$ because
the poles and the zeros satisfy the same equation (\ref{bea}).
Therefore, there are $2^{N-1}$ possible choices to form a solution
of $\Lambda(\lambda)$. With the $Z_2$ symmetry of the system, we
demonstrate that the solutions are complete.

The ground state is formed by filling all roots along the $-i\pi/4$
line (as shown in Fig.1(a)) and the ground state energy reads
$E_g=-2\cot\frac{\pi}{2N}$ which is slightly different from that of
the periodic boundary condition case. The elementary excitations of
the system can be constructed by digging some holes in the lower
solution line and putting the same number of particles in the upper
solution line (as shown in Fig.1(b)). However, the positions of the
holes and the particles are not arbitrary but obey the selection
rules of $z_j\neq z_k\pm i\pi/2$. That means if there is a hole at
$-k$, there must be a particle at $k$ (as shown in Fig.2). The
energy of a particle-hole excitation is thus $\epsilon(k)=4\sin|k|$.
Such an excitation character is quite unlike  to that in the usual
Luttinger liquids, where both the forward scattering and backward
scattering are allowed and there is no constrain for the
particle-hole excitations besides the Pauli principle in the charge
neutral sector. In the present topological boundary case, each
particle with momentum $k$ must lock a hole with momentum $-k$ to
form a virtual bound state, indicating the topological nature of the
excitations.

In conclusion, we developed a general method for diagonalizing the
integrable models without $U(1)$ symmetry. As an example, we
constructed the exact solution of the $XXZ$ spin-ring with
topological boundary condition. We remark that the present method
could be used to other integrable models without $U(1)$ symmetry.
For those models, some off-diagonal elements of the monodromy matrix
enter into the transfer matrix expression $\tau(\lambda)$. With the
commutation relations derived from the corresponding Yang-Baxter
equation, similar relation of
$\Lambda(\theta_j)\Lambda(\theta_j-\eta)\sim\Delta_q(\theta_j)$  can
be obtained from some operator identities, with which a
modified $T-Q$ relation as well as the usual BAEs can be
constructed. Details will be given elsewhere.

Y.Wang and J. Cao thank F. Gong, Y. Liu, G.M. Zhang and X.W. Guan
for valuable discussions. W. Yang and K. Shi acknowledge the
hospitality of IoP, CAS during their visit. The authors acknowledge
Y.Z. Jiang's help for numerical simulations. This work is supported
by the NSF of China, the National Program for Basic Research of MOST
(973 project).

*Corresponding author: yupeng@iphy.ac.cn


\begin{thebibliography}{99}

\bibitem{yang2} C. N. Yang, Phys. Rev. Lett. {\bf 19}, 1312 (1967); Phys. Rev. {\bf 168}, 1920
(1968).
\bibitem{bax1} R. J. Baxter, Ann. Phys. (N.Y.) {\bf 70}, 323 (1972).
\bibitem{bax2} R. J. Baxter, {\it Exactly Solved Models in Statistical
Mechanics} (Academic Press, London, 1982).
\bibitem{sk} E. K. Sklyanin and L. D. Faddeev, Sov. Phys. Dokl. {\bf 23}, 902
(1978).
\bibitem{xyz2} L. A. Takhtadzhan and L. D. Faddeev, Rush. Math.
Surveys {\bf 34}, 11 (1979).
\bibitem{yb}For details, see for example, V. E. Korepin, N. M. Boliubov, and A. G. Izergin, {\it Quantum Inverse Scattering
Method and Correlation Functions} (Cambridge University Press,
Cambridge, 1993).
\bibitem{xyz1} R. J. Baxter, Phys. Rev. Lett. {\bf 26}, 832 (1971); {\bf 26}, 834 (1971);
Ann. Phys. (N.Y.) {\bf 70}, 193 (1972).
\bibitem{G1} C.M. Yung and M.T. Batchelor, Nucl. Phys. B {\bf 446}, 461 (1995).
\bibitem{G2} M.T. Batchelor, R. J. Baxter, M. J. O'Rourke, and C. M. Yung, J. Phys. A.
Math. Gen. {\bf 28}, 2759 (1995).
\bibitem{G3} W. Galleas,  Nucl. Phys. B {\bf 790}, 524 (2008).
\bibitem{G4} H. Frahm, J. H. Grelik, A. Seel, and T. Wirth, J. Phys. A:
Math. Theor. {\bf 44}, 015001 (2011).
\bibitem{G5} S. Niekamp, T. Wirth, and H. Frahm, J. Phys. A: Math. Theor. {\bf 42}, 195008
(2009).
\bibitem{G6} G. Niccoli, Nucl. Phys. B {\bf 870}, 397 (2013).
\bibitem{G7} G. Niccoli, J. Phys. A: Math. Theor. {\bf 46}, 075003 (2013).
\bibitem{G8} J. Cao, H. Q. Lin, K. J. Shi, and Y. Wang, Nucl. Phys. B {\bf 663}, 487 (2003).
\bibitem{G9} R. I. Nepomechie, J. Phys. A: Math. Gen. {\bf 34}, 9993
(2001); Nucl. Phys. B {\bf 662}, 615 (2002).
\bibitem{G10} W. L. Yang, X. Chen, J. Feng, K. Hao, K. J. Shi, C. Y. Sun, Z. Y. Yang, and Y. Z. Zhang, Nucl. Phys. B {\bf 847}, 367 (2011).
\bibitem{G11} G. Niccoli, J. Stat. Mech. P10025 (2012).
\bibitem{G12} A. M. Grabinski and H. Frahm, J. Phys. A: Math. Theor. {\bf 43},
045207 (2010).
\bibitem{sk1} E. K. Sklyanin, Lect. Notes Phys. {\bf 226}, 196
(1985); J. Sov. Math. {\bf 31}, 3417 (1985).
\bibitem{sk2} E. K. Sklyanin, Prog. Theor. Phys. Suppl. {\bf 118},
35 (1995).
\bibitem{add1} C. Winkelholz, R. Fazio, F. W. J. Hekking, and Gerd
Sch\"{o}n, Phys. Rev. Lett. {\bf 77}, 3200 (1996).
\bibitem{add2} R. Fazio, F. W. J. Hekking, and A. A. Odintsov, Phys. Rev. Lett. {\bf 74},
1843 (1995).
\bibitem{add3} J.-S. Caux, H. Saleur, and F. Siano, Phys. Rev. Lett. {\bf 88},
106402 (2002).


%\bibitem{m1} A. Y. Kitaev, Phys. Usp. {\bf 44}, 131 (2001).
%\bibitem{m2} L. Fu and C. L. Kane, Phys. Rev. Lett. {\bf 100},
%096407 (2008).
%\bibitem{m3} R. M. Lutchyn, J. D. Sau and S. Das Sarma, Phys. Rev.
%Lett. {\bf 105}, 077001 (2010).
%\bibitem{m4} V. Mourik, K. Zuo, S. M. Frolov, S. R. Plissard, E. P. M. Bakkers and L. P. Kouwenhoven, Science {\bf 336}, 1003 (2012).

\end{thebibliography}
\end{document}